\documentclass[aps,prl,twocolumn,groupedaddress]{revtex4}
\usepackage[dvips]{graphics, color}

\begin{document}

\title{Lifetime of molecule-atom mixtures near a Feshbach resonance in $^{40}$K}

\author{C. A. Regal, M. Greiner, and D. S. Jin }
\thanks{Quantum Physics Division, National Institute of Standards and Technology.}
 \affiliation{JILA, National Institute of Standards
and Technology and Department of Physics, University of Colorado,
Boulder, CO 80309-0440}

\date{\today}

\begin{abstract} We report a dramatic magnetic field dependence in the lifetime of trapped,
ultracold diatomic molecules created through an s-wave Feshbach
resonance in $^{40}$K.  The molecule lifetime increases from less
than 1 ms away from the Feshbach resonance to greater than 100 ms
near resonance.  We also have measured the trapped atom lifetime
as a function of magnetic field near the Feshbach resonance; we
find that the atom loss is more pronounced on the side of the
resonance containing the molecular bound state.
\end{abstract}

\maketitle

Scattering resonances known as Feshbach resonances occur when the
collision energy of two free atoms coincides with that of a
molecular state in a closed channel. By varying the strength of an
external magnetic field experimenters can tune the relative
atom-molecule energy through the Zeeman effect. This has enabled
control over the strength of cold atom interactions, characterized
by the s-wave scattering length $a$, for both Bose and Fermi gases
\cite{Inouye1998a,Cornish2000a,Loftus2002a,Dieckmann2002a,O'Hara2002a,Bourdel2003a}.
More recently these resonances have allowed efficient creation of
ultracold, weakly bound molecules
\cite{Donley2002a,Regal2003c,Herbig2003a,Durr2003a,Cubizolles2003a,Jochim2003a,Strecker2003a}.

For Fermi gases the regime of strong atom coupling near an s-wave
Feshbach resonance is particularly interesting.  For a
sufficiently quantum degenerate, two-component Fermi gas of atoms,
it is predicted that the resonance can provide a continuous
crossover between two superfluid regimes, that of a
Bardeen-Cooper-Schreiffer (BCS) superfluid and Bose-Einstein
condensate (BEC) of strongly bound pairs
\cite{Holland2001a,Timmermans2001a,Milstein2002a,Ohashi2003a}.
However, experimental realization will require that the molecular
and atomic gases are sufficiently stable against inelastic decay.
In particular, the lifetimes should be longer than both the
collision time in the gas as well as the oscillator period of the
external trapping potential. This requirement is nontrivial
because inelastic collision rates typically increase by orders of
magnitude near a Feshbach resonance.

In this Letter we present a systematic study of the lifetime of
trapped molecules in the presence of atoms as well as the lifetime
of exclusively atoms near a Feshbach resonance. The molecules,
which are created using the Feshbach resonance, are highly
vibrationally excited. Thus, one expects large trap loss rates due
to collisional vibrational quenching of the molecules
\cite{Balakrishnan1998a,Yurovsky2000a,Soldan2002a}. Yet, very near
a Feshbach resonance the size of the molecules become extremely
large. In this regime the wavefunction of the molecules has much
less overlap with that of tightly bound molecules.  Thus, the
theoretical expectations for the lifetime of these molecules are
less clear. We show that the molecule lifetime increases
dramatically in this exotic regime.

For the case of inelastic atomic collisions, experiments using
bosons have seen dramatic enhancement of two and three-body loss
rates at Feshbach resonances
\cite{Stenger1999c,Roberts2000a,Weber2003a,Volz2003a}. However, in
general a suppression of three-body decay is expected for s-wave
interactions in a two-component Fermi gas due to Fermi statistics.
The magnitude of this suppression near a Feshbach resonance is
unknown, although theoretical progress has been made
\cite{Esry2001a,Petrov2003a,Suno2003a}. Experiments suggest that
three-body processes are not completely suppressed because strong
inelastic loss has been observed near fermionic Feshbach
resonances where two-body inelastic processes are not expected
\cite{Regal2003a,Dieckmann2002a,Bourdel2003a}. Here we observe
resonantly enhanced inelastic loss, but we find that the largest
loss rate occurs at a magnetic field that is shifted with respect
to the peak in elastic interactions.

The experiments reported here employ previously developed
techniques for cooling and spin state manipulation of a gas of
$^{40}$K atoms. Because of the quantum statistics of fermions a
mixture of two components, for example atoms in different internal
spin states, is required to have s-wave interactions in the
ultracold gas. With a total atomic spin $f=9/2$ in its lowest
hyperfine ground state, $^{40}$K has ten available Zeeman
spin-states $|f,m_f\rangle$, where $m_f$ is the spin projection
quantum number. Mixtures of atoms in two of these spin states are
used in evaporative cooling of the gas, first in a magnetic trap
and then in a far-off-resonance optical dipole trap.  The final
experiments are performed with an incoherent mixture of atoms with
$\sim 50 \%$ in each of two Zeeman levels.

\begin{figure} \begin{center}
\scalebox{.7}[.7]{\includegraphics{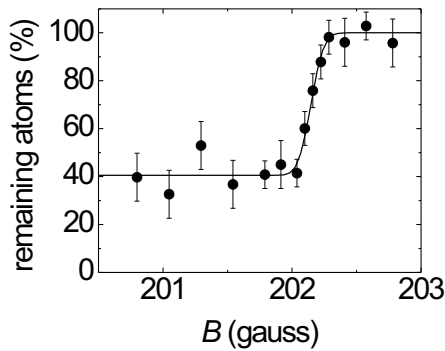}}
\caption{Remaining atom number after creating molecules by
lowering the magnetic field across the Feshbach resonance to $B$.
For this measurement we start with a two-component Fermi gas in
the $|9/2,-7/2\rangle$ and $|9/2,-9/2\rangle$ states.  The line is
a fit to an error function from which the center position is
determined to be $202.15 \pm 0.06$ G.} \label{vsB}
\end{center}
\end{figure}

We access an s-wave Feshbach resonance between atoms in the
$|9/2,-7/2\rangle$ and $|9/2,-9/2\rangle$ states located at a
magnetic field $B$ of $\sim 202$ G
\cite{Bohn2000a,Loftus2002a,Regal2003a,pwave}.  To create
molecules we adiabatically ramp the magnetic field across the
Feshbach resonance.  The magnetic field is then lowered from $B =
207$ G at a rate between 0.5 G/ms and 1.5 G/ms to various final
magnetic field values on the repulsive side of the resonance. The
number of atoms remaining following the sweep is determined from
an absorption image of the expanded atom cloud at a magnetic field
below and away from the Feshbach resonance. Since the light for
these images is resonant with an atomic transition, but not any
molecular transitions, we selectively detect only atoms. Similar
to Ref. \cite{Regal2003c}, which used a different $^{40}$K
Feshbach resonance, atoms disappear in a narrow magnetic field
region near the Feshbach resonance (Fig. \ref{vsB}). We interpret
the missing fraction of atoms as diatomic molecules as
demonstrated in \cite{Regal2003c}.

\begin{figure} \begin{center}
\scalebox{.7}[.7]{\includegraphics{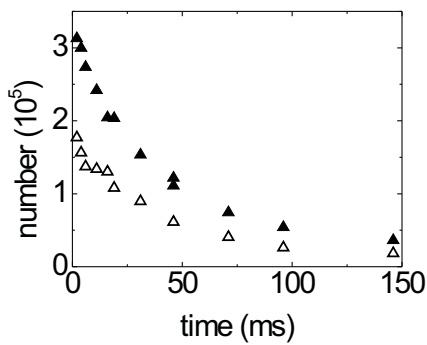}}
\caption{Remaining total atom number without (open triangles) and
with (solid triangles) transferring the molecules back to atoms.
These data are plotted as a function of the amount of time spent
at a magnetic field of $B=201.48$ G ($\Delta B$ = -0.6 G). The
curves are subtracted to determine the molecule decay rate.}
\label{lifetime}
\end{center}
\end{figure}

We can measure the number of molecules present in the sample by
converting the molecules back to atoms with one of two techniques
\cite{Regal2003c}. We can apply radio frequency waves near the
$|9/2,-7/2\rangle$ $\rightarrow$ $|9/2,-5/2\rangle$ atomic
transition during time-of-flight expansion to dissociate the
molecules into free atoms in the $|9/2,-9/2\rangle$ and
$|9/2,-5/2\rangle$ states \cite{rfpower}. We can then detect only
the molecules by selectively imaging the atom population in the
initially unoccupied $|9/2,-5/2 \rangle$ state. Alternatively, we
can ramp the magnetic field back across the Feshbach resonance to
convert the molecules back into atoms. We then measure the
resulting total number of atoms with absorption imaging. In this
case, to determine the molecule lifetime we must subtract a
measurement of the atom lifetime.  Figure \ref{lifetime} shows a
typical measurement of the atom and molecule decay. The molecule
decay rate is determined by subtracting the two curves and
extracting the initial decay rate.

The first set of lifetime measurements presented in this Letter
was performed after evaporating the Fermi gas by lowering the
optical trap to a relatively small depth. The final trap was
characterized by a radial frequency $\nu_r$ between 220 Hz and 240
Hz and an aspect ratio $\nu_r/\nu_z$ fixed at 60.   The
temperature of the gas was determined from Thomas-Fermi surface
fits to absorption images of expanded gases at magnetic fields
away from the Feshbach resonance. The typical initial cloud
temperature for these measurements was $70$ nK, with a
corresponding quantum degeneracy $T/T_F$ = $0.22$, and a total
peak density of $n_{pk}=1.5 \times 10^{13}$ cm$^{-3}$
\cite{density}.

\begin{figure} \begin{center}
\scalebox{1.3}[1.3]{\includegraphics{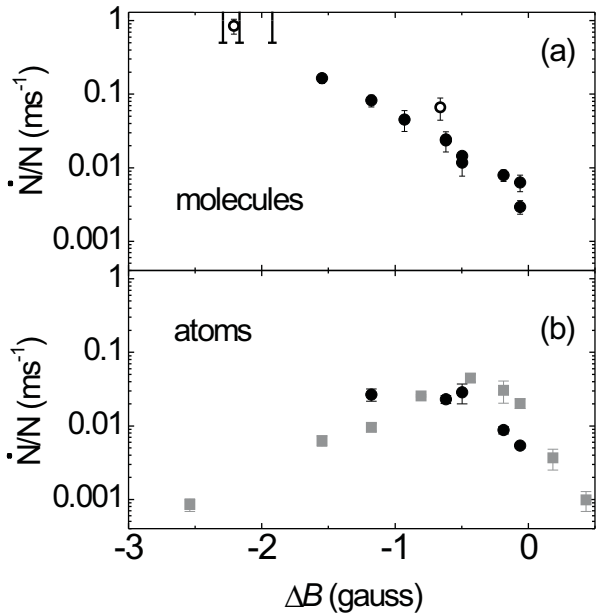}} \caption{(a)
Decay rate of molecules near a Feshbach resonance. These data are
plotted as a function of magnetic field detuning from the
$|9/2,-7/2\rangle$, $|9/2,-9/2\rangle$ Feshbach resonance peak
(solid circles) and $|9/2,-5/2\rangle$, $|9/2,-9/2\rangle$
Feshbach resonance peak (open circles). The bars indicate a lower
limit on the decay rate of $|9/2,-7/2\rangle$, $|9/2,-9/2\rangle$
molecules.  The fraction of atoms present for these measurements
ranged from 0.4 to 0.6, as indicated in Fig. {\protect \ref{vsB}}.
(b) Decay rate of atoms in the $|9/2,-7/2\rangle$,
$|9/2,-9/2\rangle$ states. The black circles are data with
molecules present; the grey squares are data taken with no
deliberate creation of molecules.  All of these measurements start
with the Fermi gas in a relatively shallow optical trapping
potential at a typical temperature of 70 nK and a typical density
of $n_{pk}=1.5 \times 10^{13}$ cm$^{-3}$.} \label{molecules}
\end{center}
\end{figure}

\begin{figure} \begin{center}
\scalebox{1.5}[1.5]{\includegraphics{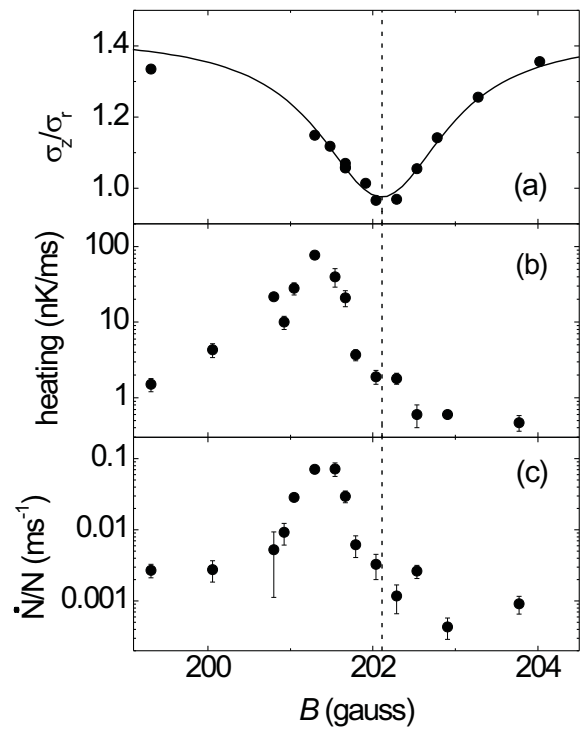}} \caption{(a)
Anisotropic expansion of a nearly equal mixture of the
$|9/2,-7/2\rangle$ and $|9/2,-9/2\rangle$ states of $^{40}$K near
a Feshbach resonance.  The aspect ratio of the cloud
$\sigma_z/\sigma_r$ decreases at the Feshbach resonance peak due
to an enhanced elastic collision rate. The line is a Lorentzian
fit of the data to determine the resonance position, $B_0=202.1$
G. (b) Heating of the $|9/2,-7/2\rangle$ and $|9/2,-9/2\rangle$
gas near the Feshbach resonance. The typical initial temperature
is 670 nK. (c) Loss of the $|9/2,-7/2\rangle$ and
$|9/2,-9/2\rangle$ gas near the Feshbach resonance.  The typical
initial density of the cloud is $n_{pk}=1.4 \times 10^{13}$
cm$^{-3}$. The dotted line corresponds to the Feshbach resonance
peak ($\Delta B = 0$).} \label{atoms}
\end{center}
\end{figure}

Figure \ref{molecules} displays loss rate measurements as a
function of the detuning from the Feshbach resonance, $\Delta B =
B-B_0$.  The rates are plotted as the decay rate of the number of
molecules (atoms) $\dot{N}$ normalized to the initial molecule
(atom) number $N$.  Figure \ref{molecules}(a) shows the loss rate
of the molecules for the $|9/2,-7/2\rangle$, $|9/2,-9/2\rangle$
Feshbach resonance (solid circles), where $B_0 = 202.1$ G (see
Fig. \ref{atoms}).  In this figure we also include two
measurements of the decay rate of molecules created from the
$|9/2,-5/2\rangle$ and $|9/2,-9/2\rangle$ Feshbach resonance (open
circles), where $B_0 = 224.21$ G \cite{Regal2003b}. The solid
circles in Fig. \ref{molecules}(b) show the corresponding atom
loss rate with molecules present.

We find that the lifetime of the molecules increases dramatically
as the Feshbach resonance is approached. Away from the Feshbach
resonance the molecule lifetime is $\sim 1$ ms as reported in Ref.
\cite{Regal2003c}. Near the resonance the lifetime is greater than
100 ms, similar to that of atoms.  It is interesting to compare
the size of the molecules to the interatomic spacing of the atoms,
which in the peak density region of the initial atom cloud is
$\sim 7000$ $a_0$. The molecule size as given by $a/2$ becomes
$7000$ $a_0$ at $\Delta B \approx - 0.1$ G \cite{width}. This
suggests that very close to the Feshbach resonance the system
exists in an exotic regime where atoms are likely to lie within
the extent of a molecule.

We have also performed experiments in which we do not deliberately
create molecules; in this way we can measure exclusively the atom
lifetime. Instead of ramping the magnetic field to reach the
Feshbach resonance, we access the Feshbach resonance by suddenly
changing the Zeeman level of one of the two spins in the system.
(This will result in a nonequilibrium sample.) The atom loss rate
is then measured by watching the decay of the atom number in time
to determine the initial atom loss rate. The results are shown as
the squares in Fig. \ref{molecules}(b). In general, we find that
the atom lifetime is similar whether or not there is a large
molecule component in the gas.

Using the spin-changing technique described above we have also
measured the loss rate in a trap whose depth is large compared to
the atoms' energy. In this case the loss rate is more accurately
determined, since heating does not lead to evaporative loss. For
these data the optical trapping potential was recompressed after
evaporative cooling. The final trap had a depth of $\sim 10$
$\mu$K and a radial trapping frequency of $\nu_r = 725$ Hz. The
final gas was at $T=670$ nK and $T/T_F = 0.85$, but had a similar
peak density to data in Fig. \ref{molecules}, $n_{pk}=1.4 \times
10^{13}$ cm$^{-3}$. To determine the atom loss and heating rates
we measure the total atom number as a function of time until $\sim
20 \%$ of the atoms are lost. From the slope of these data in time
we determine the heating and loss rates (Fig. \ref{atoms}(b),(c)).

Additionally, in Fig. \ref{atoms} we plot hydrodynamic expansion
data from which we determine the position of the Feshbach
resonance. To measure the aspect ratio of the expanded cloud
$\sigma_z/\sigma_r$ we started with a gas in a trap characterized
by $\nu_r=440$ Hz at $T/T_F=0.55$ and $n_{pk}=1.7 \times 10^{13}$
cm$^{-3}$. We access the Feshbach resonance with the previously
described spin changing technique.  We then turn off the optical
trapping light, and the gas expands for a total of 21.3 ms, where
7 ms of this expansion takes place at the magnetic field value
near the Feshbach resonance. As shown in Refs.
\cite{Regal2003b,Bourdel2003a,Gehm2003a,Gupta2003a} anisotropic
expansion of the gas in this regime is a signature of a large
elastic collision rate. Thus, we interpret the magnetic field
location of the maximal decrease in the aspect ratio of the
expanded cloud as the position of the Feshbach resonance, $B_0$.
We find that $B_0 = 202.1 \pm 0.1$ G (Fig. \ref{atoms}(a)).

As shown in Fig. \ref{atoms}, we find large loss rates on
primarily the repulsive side of the $|9/2,-7/2\rangle$,
$|9/2,-9/2\rangle$ Feshbach resonance peak. Recognize that the
$|9/2,-7/2\rangle$ and $|9/2,-9/2\rangle$ spin states are
distinguished as being the two lowest energy states of the
$^{40}$K atom at these magnetic fields; they are therefore immune
to any two-body losses associated with the s-wave resonance
\cite{Bohn2000a}. Hence, we interpret the loss as three-body
recombination associated with the Feshbach resonance.  The strong
heating can be explained by the density dependence of the
three-body loss process as well as by the binding energy released
in a recombination process to a weakly bound molecular state
\cite{Weber2003a}. We note that the observed peak in the inelastic
processes for an s-wave Feshbach resonance in $^6$Li is also
shifted to the repulsive side of the maximum of the elastic
scattering \cite{Bourdel2003a,Dieckmann2002a}.

In conclusion, we have measured the magnetic-field dependence of
the lifetime of both molecules and atoms near an s-wave $^{40}$K
Feshbach resonance. Away from resonance the molecules have a short
lifetime due to vibrational quenching, while the atoms have a long
lifetime due to Fermi statistics. However, the situation is
drastically changed very close to the Feshbach resonance as the
molecule size begins to be comparable to the interatomic spacing.
Here the molecule decay rate is suppressed by orders of magnitude,
while the atoms exhibit strong three-body recombination rates. The
atom loss occurs primarily on the repulsive side of the resonance
where the weakly bound molecular state exists.  We speculate that
this state plays an important role in the asymmetry of the
three-body recombination. Furthermore, our results suggest that
near a Feshbach resonance the strong overlap of atom pair and
molecule wavefunctions plays an important role in inelastic
processes.

This study shows that the outlook is good for studying long-lived
molecules and strongly interacting Fermi atoms.  The molecule
lifetime near the Feshbach resonance has been shown to be much
longer than collision times and trap oscillator periods (Fig.
\ref{molecules}). Further, the attractive (high magnetic field)
side of the resonance, where fermionic superfluidity is predicted,
displays strong elastic interactions with a relatively
inconsequential inelastic rates (Fig. \ref{atoms}).

We thank C. E. Wieman, E. A. Cornell, and M. Holland for useful
discussions and J. Smith for experimental assistance. This work
was supported by NSF and NIST; C. A. R. acknowledges support from
the Hertz Foundation.


\begin{thebibliography}{10}
\bibitem{rfpower}
The photodissociation rf pulse is detuned by $\sim 100$ kHz beyond
the
  threshold of molecular dissociation and applied for 50 to 200 $\mu$s. The rf
  power corresponds to a $\sim 15$ $\mu$s bare atom $\pi$-pulse on resonance
  with the $|9/2,-7/2\rangle$ $\rightarrow$ $|9/2,-5/2\rangle$ transition.
\bibitem{density}
The systematic uncertainty in the density is $50\%$ due to an
uncertainty in
  the absolute number of atoms.
\end{thebibliography}

\end{document}